\pdfoutput=1

\documentclass[twoside,twocolumn,9pt]{article}
\usepackage{extsizes}
\usepackage[super,sort&compress,comma]{natbib} 
\usepackage[version=3]{mhchem}
\usepackage[left=1.5cm, right=1.5cm, top=1.785cm, bottom=2.0cm]{geometry}
\usepackage{balance}
\usepackage{widetext}
\usepackage{times,mathptmx}
\usepackage{sectsty}
\usepackage{graphicx} 
\usepackage{lastpage}
\usepackage[format=plain,justification=raggedright,singlelinecheck=false,font={stretch=1.125,small,sf},labelfont=bf,labelsep=space]{caption}
\usepackage{float}
\usepackage{fancyhdr}
\usepackage{fnpos}
\usepackage[english]{babel}
\usepackage{array}
\usepackage{droidsans}
\usepackage{charter}
\usepackage[T1]{fontenc}
\usepackage[usenames,dvipsnames]{xcolor}
\usepackage{setspace}
\usepackage[compact]{titlesec}

\usepackage{gensymb}
\usepackage{units}
\usepackage{flushend}

\definecolor{cream}{RGB}{222,217,201}

\begin{document}

\pagestyle{fancy}
\thispagestyle{plain}
\fancypagestyle{plain}{

\renewcommand{\headrulewidth}{0pt}
}

\makeFNbottom
\makeatletter
\renewcommand\LARGE{\@setfontsize\LARGE{15pt}{17}}
\renewcommand\Large{\@setfontsize\Large{12pt}{14}}
\renewcommand\large{\@setfontsize\large{10pt}{12}}
\renewcommand\footnotesize{\@setfontsize\footnotesize{7pt}{10}}
\makeatother

\renewcommand{\thefootnote}{\fnsymbol{footnote}}
\renewcommand\footnoterule{\vspace*{1pt}%
\color{cream}\hrule width 3.5in height 0.4pt \color{black}\vspace*{5pt}} 
\setcounter{secnumdepth}{5}

\makeatletter 
\renewcommand\@biblabel[1]{#1}            
\renewcommand\@makefntext[1]%
{\noindent\makebox[0pt][r]{\@thefnmark\,}#1}
\makeatother 
\renewcommand{\figurename}{\small{Fig.}~}
\sectionfont{\sffamily\Large}
\subsectionfont{\normalsize}
\subsubsectionfont{\bf}
\setstretch{1.125} 
\setlength{\skip\footins}{0.8cm}
\setlength{\footnotesep}{0.25cm}
\setlength{\jot}{10pt}
\titlespacing*{\section}{0pt}{4pt}{4pt}
\titlespacing*{\subsection}{0pt}{15pt}{1pt}

\renewcommand{\headrulewidth}{0pt} 
\renewcommand{\footrulewidth}{0pt}
\setlength{\arrayrulewidth}{1pt}
\setlength{\columnsep}{6.5mm}
\setlength\bibsep{1pt}

\makeatletter 
\newlength{\figrulesep} 
\setlength{\figrulesep}{0.5\textfloatsep} 

\newcommand{\topfigrule}{\vspace*{-1pt}%
\noindent{\color{cream}\rule[-\figrulesep]{\columnwidth}{1.5pt}} }

\newcommand{\botfigrule}{\vspace*{-2pt}%
\noindent{\color{cream}\rule[\figrulesep]{\columnwidth}{1.5pt}} }

\newcommand{\dblfigrule}{\vspace*{-1pt}%
\noindent{\color{cream}\rule[-\figrulesep]{\textwidth}{1.5pt}} }

\makeatother

\twocolumn[
  \begin{@twocolumnfalse}
\vspace{3cm}
\sffamily
\begin{tabular}{m{4.5cm} p{13.5cm} }

\includegraphics{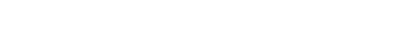} & \noindent\LARGE{\textbf{Optimizing Packing Fraction in Granular Media Composed of Overlapping Spheres$^\dag$}} \\
\vspace{0.3cm} & \vspace{0.3cm} \\

 & \noindent\large{Leah K. Roth$^{\ast}$ and Heinrich M. Jaeger} \\

\includegraphics{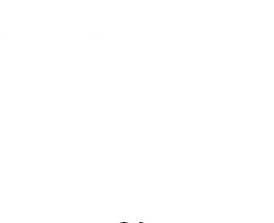} & \noindent\normalsize{What particle shape will generate the highest packing fraction when randomly poured into a container? In order to explore and navigate the enormous search space efficiently, we pair molecular dynamics simulations with artificial evolution. Arbitrary particle shape is represented by a set of overlapping spheres of varying diameter, enabling us to approximate smooth surfaces with a resolution proportional to the number of spheres included. We discover a family of planar triangular particles, whose packing fraction of $\phi \sim$ 0.73 outpaces almost all reported experimental results for random packings of frictionless particles. We investigate how $\phi$ depends on the arrangement of spheres comprising an individual particle and on the smoothness of the surface. We validate the simulations with experiments using 3D-printed copies of the simplest member of the family, a planar particle consisting of three overlapping spheres with identical radius. Direct experimental comparison with 3D-printed aspherical ellipsoids demonstrates that the triangular particles pack exceedingly well not only in the limit of large system size but also when confined to small containers.}\\

\end{tabular}

 \end{@twocolumnfalse} \vspace{0.6cm}

  ]

\renewcommand*\rmdefault{bch}\normalfont\upshape
\rmfamily
\section*{}
\vspace{-1cm}


\footnotetext{\textit{$^{a}$~James Franck Institute and Department of Physics, The University of Chicago, Chicago, IL 60637, USA. E-mail: rothl@uchicago.edu}}






The relationship between particle shape and packing density for random particle arrangements continues to be a topic of considerable interest.\cite{Bau:14,Jae:15,Tor:10} Going beyond spherical particles, there has been much recent progress for polyhedral or polygonal shapes,\cite{Haj:09,Jao:10,Bak:10,Neu:13,Ngu:15,Dam:12} ellipsoids, cuboids, or `superballs,'\cite{Don:07,Sch:10,Del:10,She:12,Tor:12,Ni:12} cylinders, cones, and frustums of different aspect ratios,\cite{Wou:07,Wou:09,Zha:11} as well as various types of particles constructed by joining disks or spheres.\cite{Sch:10,Kod:09,Sai:11,Lud:12,Phi:12,Aze:13,Mis:13,Mis:14} Furthermore, in the last few years, increasing attention has been paid to particles that are highly non-convex or are sufficiently flexible to assume non-convex shapes during the packing process.\cite{Tor:12,Rem:08,Zou:09,Gal:09,Mal:09,Lop:11,Gra:11,Bro:12,Gra:12,Men:12} In almost all cases, these studies proceeded from a given particle type to find the packing density. What has remained a major challenge is a general and systematic approach to the inverse problem: taking desired packing properties as a starting point to identify the appropriate particle shape.

To make progress, it is necessary to address three main obstacles. Firstly, shape is an infinitely variable parameter, rendering an exhaustive investigation of all possible particle shapes infeasible and instead requiring a smart approach to quickly narrow the search. The second obstacle relates to the fundamental nature of amorphous, jammed aggregates. Because a jammed system exists far from equilibrium, the packing density $\phi$ can be affected not only by particle geometry but also by boundary and processing conditions. These effects become particularly pronounced when extrapolating  from finite size experiments and simulations to the properties of an infinite system. Finally, different shapes may generate similar packing fractions when assembled into an aggregate under particular processing conditions, so that for a given level of approximation to the desired target density a multitude of shapes may emerge as viable candidates.

Here we address the specific question of finding particle shapes that achieve high random packing density.  We tackle the above obstacles by using an approach developed recently by our group,\cite{Mis:14,Mis:13} which treats the inverse problem as an optimization task within the context of artificial evolution. With this method, we turn the complex relationship between particle shape and packing density into a high-dimensional search landscape, where a physically accurate simulation generates a random packing and measures $\phi$ for a trial particle shape to be tested, and an efficient evolution algorithm mutates and updates this shape to maximize $\phi$.

\begin{figure}[h]
\centering
  \includegraphics[height=7cm]{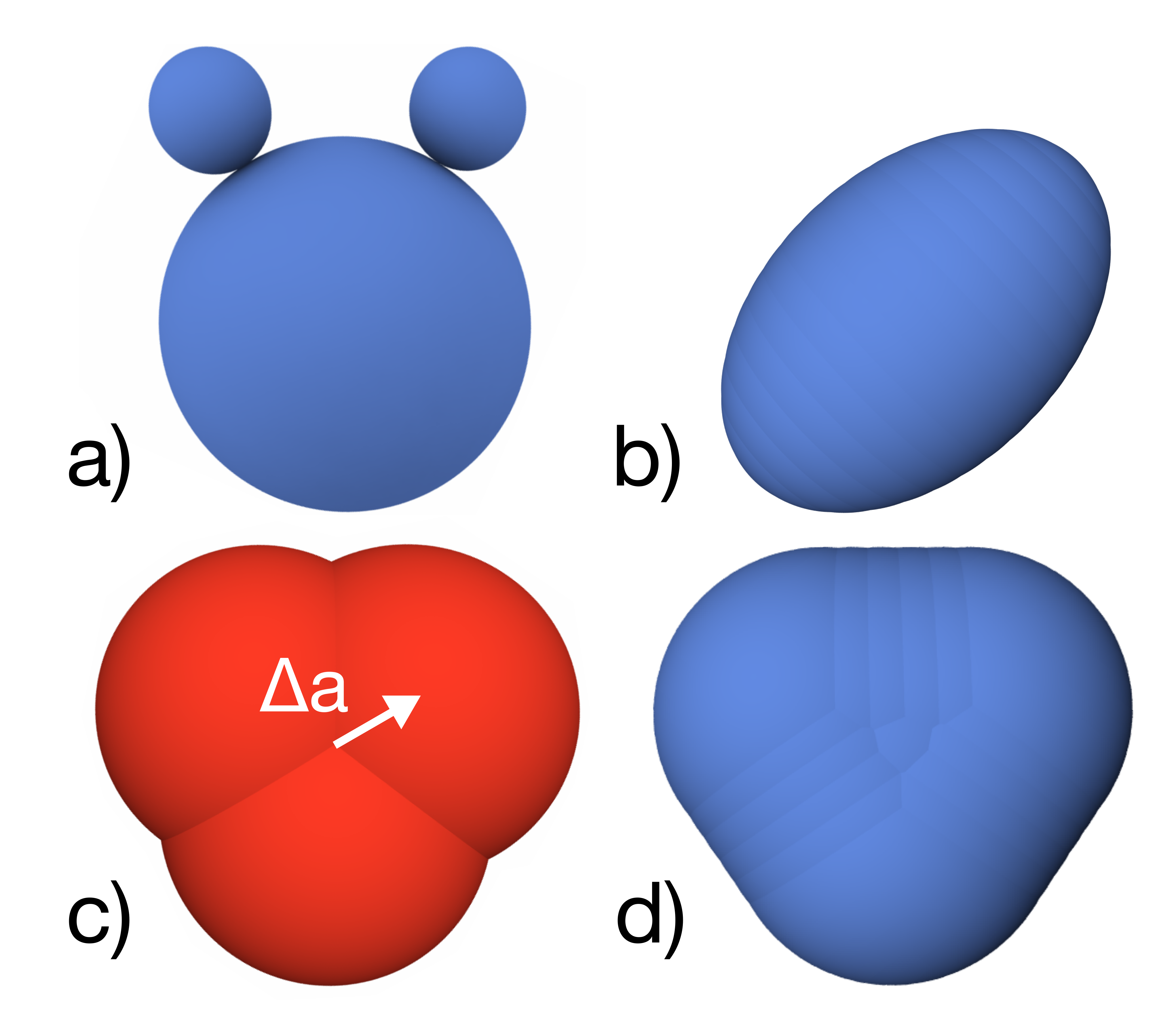}
  \caption{Particle shapes producing high random packing densities. \textit{a)} Optimized trimer for non-overlapping spheres.\cite{Mis:14,Mis:13} \textit{b)} Ellipsoid, composed of 17 overlapping spheres. \textit{c)} Overlapping-sphere trimer, representative of a class of planar triangular shapes identified by our algorithm as packing most densely. Each constituent sphere of radius $a$ is a separated by a distance $\Delta a$ from the particle's center of mass. \textit{d)} Overlapping-sphere trimer with additional spheres smoothing out the shape. The resulting packing fraction remains within the simulation uncertainty of the trimer.}
  \label{diff_shapes}
\end{figure}

In our previous work,\cite{Mis:14,Mis:13} particles of arbitrary shape were represented by a series of bonded spheres of varying radii, constrained to be touching at their contacting surfaces. With this shape representation, the particle type  found by the optimizer to produce the densest random packing was a 3-sphere `granular molecule' giving $\phi$ = 0.67 $\pm$ 0.03.  In this trimer particle, a central sphere is bonded to two smaller spheres of \nicefrac{1}{3} the diameter at a bond angle of 70\degree (Fig. \ref{diff_shapes}a). Approximating arbitrary shapes by bonded spheres is efficient in simulation due to the simplicity of force calculations when two particles come into contact. However, using non-overlapping spheres causes the surface of the particle to be corrugated. This `geometric friction' is known to affect the packing density\cite{Sch:10} and frustrates attempts to extrapolate results to particle shapes with smooth, non-corrugated surfaces.

We move beyond this approach by allowing the constituent spheres comprising an individual particle to overlap as well as vary in diameter. The spheres composing a particle are placed in sequence, so that each sphere is described by a bearing relative to the previous sphere placed, consisting of two angles, and a radius, identical to our previous `bonded sphere' parameterization. Now we introduce an additional degree of freedom, adding a parameter to specify the degree of overlap of a sphere with the previous sphere placed. With this we gain the ability to approximate a smooth, continuous surface with a precision proportional to the number of constituent spheres, while retaining the simplicity of sphere-to-sphere force calculations. In this way, we vastly expand the range of particle shapes accessible to the optimizer, enabling it to explore shapes such as smooth ellipsoids (Fig. \ref{diff_shapes}b) that previously were out of reach.

With this framework in place, we focus on the task of finding the highest packing fraction achievable with compound particles composed of $N$ overlapping, frictionless spheres of arbitrary diameters, poured into a container under gravity.  We chose this problem because it allows us to compare with shapes previously identified in the literature as good packers.  It also enables us to perform direct validation experiments with 3D-printed versions of the shapes identified by the optimizer, linking our work to a wide range of applications in the area of granular materials.

We had expected that the shape packing most densely may approximate an ellipsoid, following trends identified by Donev et al.\cite{Don:04}, or alternatively may turn out to be a variation on the trimer in Fig. \ref{diff_shapes}a. Instead, the optimizer identified a completely different shape: a planar triangular `molecule' that in its simplest form is composed of three overlapping spheres of the same diameter, arranged in a roughly equilaterial triangle  (Fig. \ref{diff_shapes}c). When additional spheres are added to smooth out dimples  (Fig. \ref{diff_shapes}d) the packing fraction is affected very little.  Both simulations and experiments indicate that this triangular particle type packs at remarkably high densities, near $\phi$ = 0.73, in the limit of infinite system size.

\begin{figure*}
\centering
  \includegraphics[height=4.7cm]{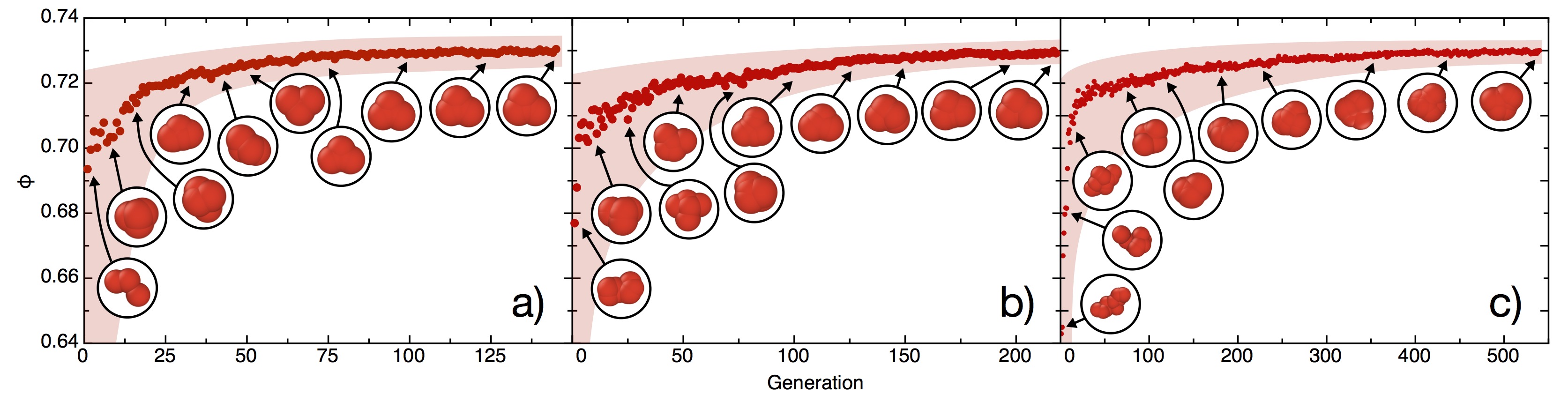}
  \caption{Evolution of the particle shape. The plots give the median packing density in each generation as the optimizer explores shape space for different number of constituent spheres $N$ = 5, 10, 25 (panels \textit{a}-\textit{c}). The images show representative snapshots of members of the shape population at various generations.}
  \label{3_gen_plots}
\end{figure*}

\section{Simulation \& Optimization}

The backbone of our optimization pipeline was the Covariance Matrix Adaptation Evolution Strategy (CMA-ES),\cite{Han:03} a robust, fast and efficient algorithm that is successful even when confronted with complex nonlinear problems.\cite{Han:10} In prior work we found CMA-ES to deliver excellent performance in optimization problems ranging from the packing of granular materials\cite{Mis:13,Mis:14} to directed self-assembly of copolymers.\cite{Qin:13:1,Qin:13:2} The evolutionary algorithm worked in concert with molecular dynamics simulations of `virtual experiments,' providing input parameters to these simulations and updating the parameters according to the simulated outcome in relation to the optimization target (for details see Ref. 25).

We used the molecular dynamics simulation platform LAMMPS\cite{Pli:95} to pour  $\sim$900 frictionless particles under gravity into a three-dimensional box of side length $\sim$10$d$, where $d$ is the characteristic size of a single particle. To mitigate wall effects, the box was periodic in $x$ and $y$, and we excluded all particles within 3$d$ from the floor of the container from our calculation of packing fraction. The forces between contacting spheres were calculated using a frictionless Hertzian model, where the material density and elastic constant of the particles were chosen to be consistent with the hard resin (Vero White, with measured Young's modulus $\approx$ 1.3GPa\cite{Mis:13}) used in our Connex Objet 350 3D-printer for experimental validation (discussed further below).

During each generation of the evolutionary algorithm, a population of 40 different particle shapes, each composed of $N$ overlapping spheres, was generated and 40 packing experiments, one for each particle type, were simulated in parallel. With this protocol, we simulated a poured aggregate for each particle shape in a population and calculated the resulting packing densities. These data were then used by the optimizer to produce the next generation of particle shapes, suitably mutated to explore shape space in search of potentially higher packing densities. Note that the optimizer could set the radii $a_i$ of one or more constituent spheres to zero; thus $N$ should be thought of as the maximum possible number of spheres per particle.

Over the course of 200---300 generations, the optimizer converged on an asymptotic solution.

\subsection{Results}

In the results reported here, we performed optimizations where we allowed the optimizer to use up to $N$ = 5, 10 or 25 spheres of arbitrary radius in composing each particle, testing a population of 40 different particle shapes in each generation. As the number of constituent spheres increases, and with it the number of free parameters, the number of generations required to converge on a solution also grows.

When only 5 spheres are used to compose each particle (Fig. \ref{3_gen_plots}a), the optimizer identified an overlapping-sphere trimer in which three spheres are arranged, in a plane, symmetrically about the particle's center of mass (Fig. \ref{diff_shapes}c). In this shape, the three spheres have the same radii $a_i$ = $a$ within our numerical accuracy, and they overlap by nearly the same amounts such that the distance from the center of mass $\Delta a$ ranges between 0.43$a$ and 0.48$a$ for each of the spheres. Our simulations predict that this shape will pack at approximately 0.729$\pm$0.003, significantly higher than the non-overlapping-sphere trimer (Fig. \ref{diff_shapes}a), and nearing theoretical predictions\cite{Bau:13} of the density of lens-shaped particles, at 0.736, as well as previous experimental results\cite{Man:05} for specific aspherical ellipsoids at 0.735 to 0.74.

As we increase the number of allowed spheres used to compose each particle from 5 to 10 (Fig. \ref{3_gen_plots}b), the final result of the optimization retains the general characteristics of the trimer, but with some variations. It seems that the 10-sphere optimization attempts to smooth the slightly scalloped edge of the trimer while remaining planar and overall triangular. Within our numerical accuracy, the final packing fraction is indistinguishable from that for $N$ = 5.

Allowing the optimizer to choose the positions of up to $N$ = 25 spheres continues this trend  (Fig. \ref{3_gen_plots}c). The final result moves further from the scalloped, 3-sphere shape and more towards a smooth triangular shape. Indeed, if we smooth out the edges of the trimer by adding additional spheres `by hand' (Fig. \ref{diff_shapes}d), the packing fraction increases only slightly to 0.733$\pm$0.005, within error of the trimer result. However, while the optimizer tends to prefer the smoother shape given the additional freedom of more spheres, it eventually runs out of steam when the uncertainty in packing density determination at each generation becomes larger than the potential advantage gained from smoothing out the triangular shape. Interestingly, the optimizer retains the planar shape independent of the number of spheres allowed for each particle, indicating that the oblate nature of the trimer plays a definite role in its success.

\begin{figure}[h]
\centering
  \includegraphics[height=5.5cm]{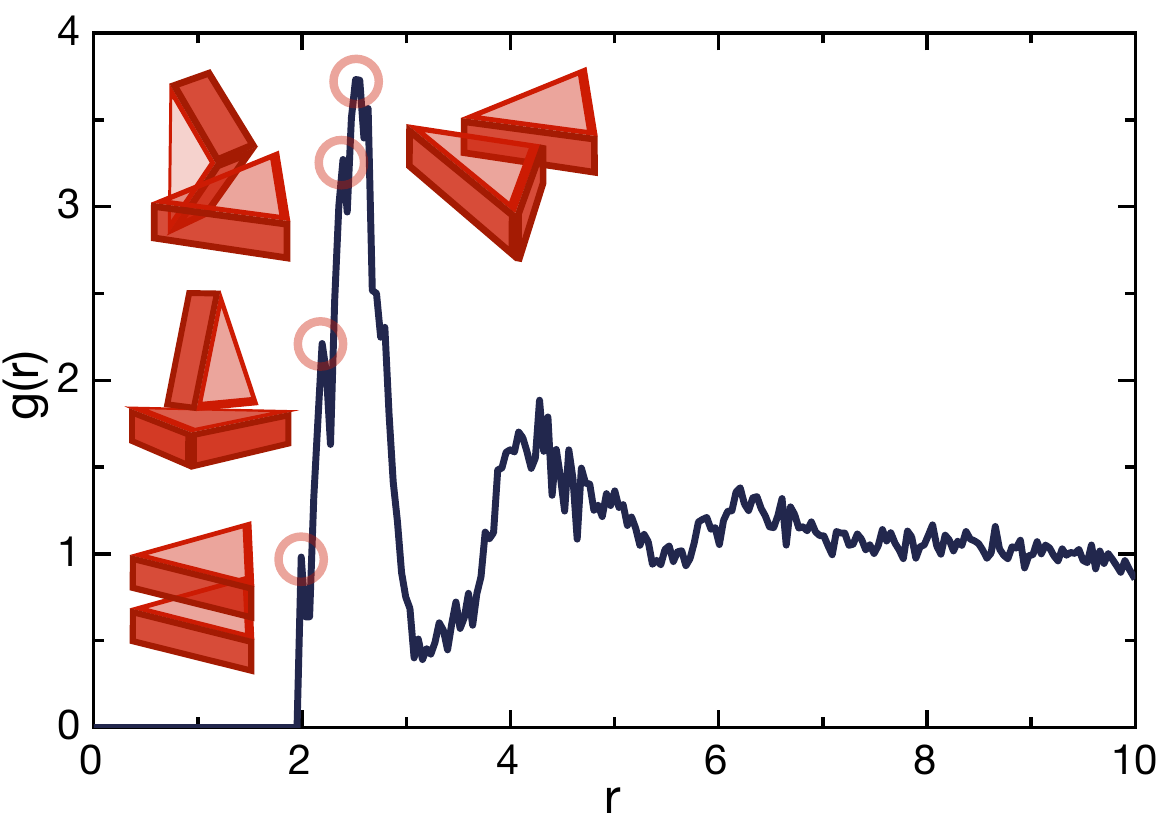}
  \caption{Radial distribution function $g$($r$) for the overlapping-sphere trimer as function of center-to-center distance $r$ between trimers. The radius of each individual sphere making up a trimer is $a$, so that two particles can touch at $r$ = $2a$. Common configurations of two contacting trimers corresponding to four features in $g$($r$) are indicated with stylized triangles for clarity.}
  \label{g_of_r}
\end{figure}

\begin{figure*}
\centering
  \includegraphics[height=6cm]{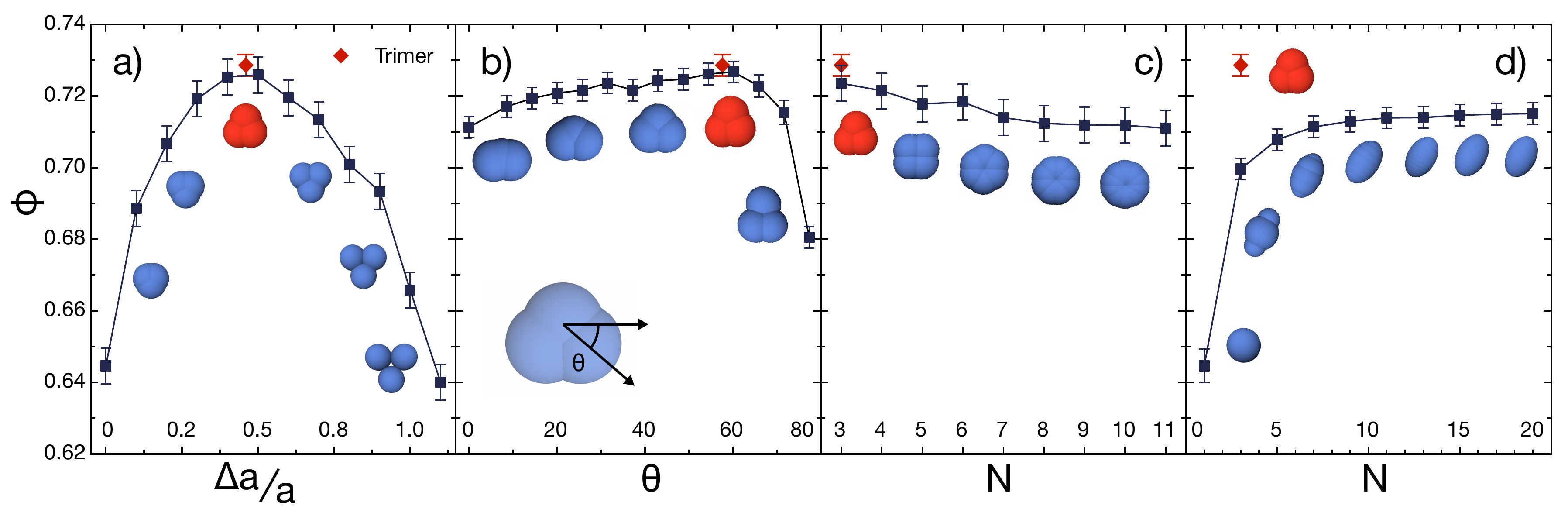}
  \caption{\textit{a)} Increasing or decreasing the center-to-center distance $\nicefrac{\Delta a}{a}$ of the constituent spheres  from the  value identified by the optimizer negatively affects the packing density. \textit{b)} Dependence on in-plane sphere arrangement, parameterized by angle $\theta$, as defined in the inset. \textit{c)} Adding spheres in-plane to the trimer does not positively affect its packing properties. \textit{d)} Packing density of prolate ellipsoids approximated by $N$ overlapping spheres.}
  \label{shape_plots}
\end{figure*}

In what follows we focus on the trimer composed of three identical, overlapping spheres (Fig. \ref{diff_shapes}c) as the representative of a family of planar triangular shapes that generate similarly dense random packings. The simplicity of this trimer allows us to readily investigate the dependence of each sphere's position on the aggregate packing fraction, and is also an interesting testament to the success of non-convex, corrugated shapes in achieving large $\phi$. 

 We have found no prominent structural features in the packing arrangement responsible for this remarkably dense packing. Significant local particle ordering or a preference for particular orientations between neighboring particles appear to be absent. In Fig.~\ref{g_of_r}, we plot the radial distribution function for the trimer, which was calculated in a packing composed of about 10,000 particles. These data show peaks for a number of likely configurations of contacting particles, but they do not exhibit characteristics that could pinpoint the reason for this shape's success.

On the other hand, the trimer's particular geometry appears to place it near the optimal aspect ratio identified previously for three-dimensional lens-shape particles. Defining the aspect ratio $\alpha$ in the same way as Baule et al.\cite{Bau:13}, by the ratio of particle width to particle length measured along the axis of symmetry, we find $\alpha$ $\sim$ 0.74 for the trimer. For oblate, lens-shaped particles, mean-field theory predicts\cite{Bau:13}  $\phi$ to peak within the region $\alpha$ = 0.75-0.85. 

\subsection{Robustness}

It is valuable to investigate how perturbations in shape affect $\phi$. In Fig. \ref{shape_plots}, we varied the distance of each sphere from the trimer's center of mass, the angle formed by two constituent spheres and the horizontal, and the number of spheres arranged isotropically around the particle's center of mass.

From Fig.~\ref{shape_plots}a it is clear that any radial expansion or contraction of the trimer from the value optimized by the evolutionary algorithm will negatively affect the packing fraction. The quantity $\nicefrac{\Delta a}{a}$ describes the degree of separation between the spheres, where $\Delta a$ is the distance of each sphere from the center of mass of the trimer and $a$ is the radius of each constituent sphere. As the spheres move closer, $\phi$ decreases toward the random packing density of single spheres. Once $\nicefrac{\Delta a}{a}$ increases beyond 0.5, the additional space created between constituent spheres, coupled with the surface corrugation, again drives the packing fraction lower. This non-monotonicity is also seen in a similar context in Refs. 11 and 23, indicating a more general relationship between the degree of corrugation of overlapping spheres and packing fraction.

Varying the angle $\theta$ between two constituent spheres and the horizontal, while keeping the distance between the lower spheres fixed, exhibits a much less sensitive effect on packing fraction, which only decreases by a large margin when the three spheres begin to separate (Fig.~\ref{shape_plots}b). Adding additional spheres within the plane of the trimer decreases $\phi$, indicating that the triangular shape of the trimer contributes to its success over a more square or disk like shape (Fig.~\ref{shape_plots}c).

Previous research showed\cite{Don:04} that prolate ellipsoids with an aspect ratio of 1.5:1 pack at $\phi \sim$ 0.71. To probe the effect of geometric friction, we composed ellipsoids of this aspect ratio using varying numbers of spheres, $N$. As $N$ increases, the ellipsoid becomes increasingly smooth (Fig.~\ref{shape_plots}d). When about 10 spheres are used, the packing fraction asymptotes at almost exactly the value predicted, significantly below that of the overlapping-sphere trimer.

\begin{figure}[h]
\centering
  \includegraphics[height=5.5cm]{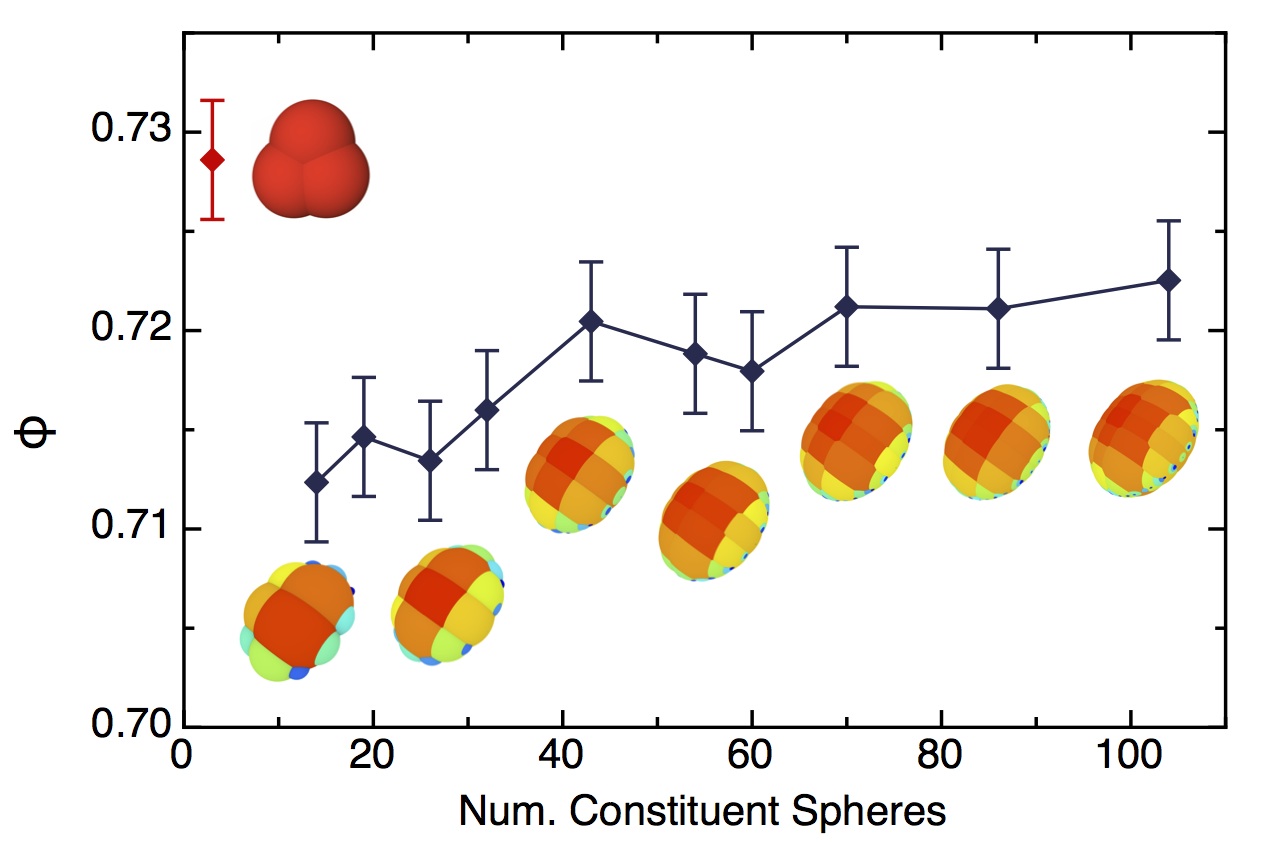}
  \caption{Packing density of aspherical ellipsoids with aspect ratio 1.25:1:0.8, approximated with increasing number $N$ of overlapping spheres.}
  \label{asph_ellipse_sim}
\end{figure}

Further work by Chaikin, Torquato and coworkers indicated\cite{Man:05,Don:04} that aspherical, non-axisymmetric ellipsoids of aspect ratio 1.25:1:0.8 generate random packings  with $\phi$ approaching 0.74. As before, we approximated this particular type of ellipsoid by increasing $N$ until the simulated packing fraction became independent of the number of spheres used (Fig. \ref{asph_ellipse_sim}). Our simulations show an asymptotic packing fraction of 0.722$\pm$0.005, still smaller than for the overlapping-sphere trimer. The discrepancy between this asymptotic $\phi$ and earlier reports \cite{Don:04} is unlikely to be caused by residual dimples in the simulated particles, since at $N$ = 104 the individual particle volume differs from that of a completely smooth ellipsoid by less than 0.7\%. We discuss other potential reasons below.

\section{Experimental Validation \& Discussion}

To verify the simulation results, we printed the overlapping trimer in hard resin using a high-resolution Objet Connex 350 3D printer. We produced $\sim$10,000 individual particles, each with a volume of 66 mm\textsuperscript{3} and each of the three constituent spheres with a radius of 2 mm. For comparison, we also we 3D-printed  $\sim$10,000 aspherical ellipsoids of aspect ratio 1.25:1:0.8, each with a volume of 32 mm\textsuperscript{3} and an average radius of 2 mm.

In the experiments, the particles were poured into cylinders to measure the resulting packing  density.  We mobilized friction by hand-tapping the packing so that the density asymptotically approached that of poured, frictionless particles.\cite{Mis:14} Though other packing protocols engender different results,\cite{Neu:13} hand-tapping is straightforward and feasible in a wide range of practical situations. We chose tapping over the use of liquid lubricant, because mixing particles with lubricant tends to introduce small bubbles that become trapped at the particle-liquid interfaces and can bias the measured packing density toward higher values.

While the simulations were performed with periodic boundary conditions in $x$ and $y$, in the experiments we measured the packing fraction in  finite-sized cylinders, introducing boundary effects at the walls. To identify and mitigate these effects, we systematically varied the cylinder diameters and volumes, and filled each cylinder to a depth sufficient to render effects from the container floor negligible.

\begin{figure}[h]
\centering
  \includegraphics[height=3.5cm]{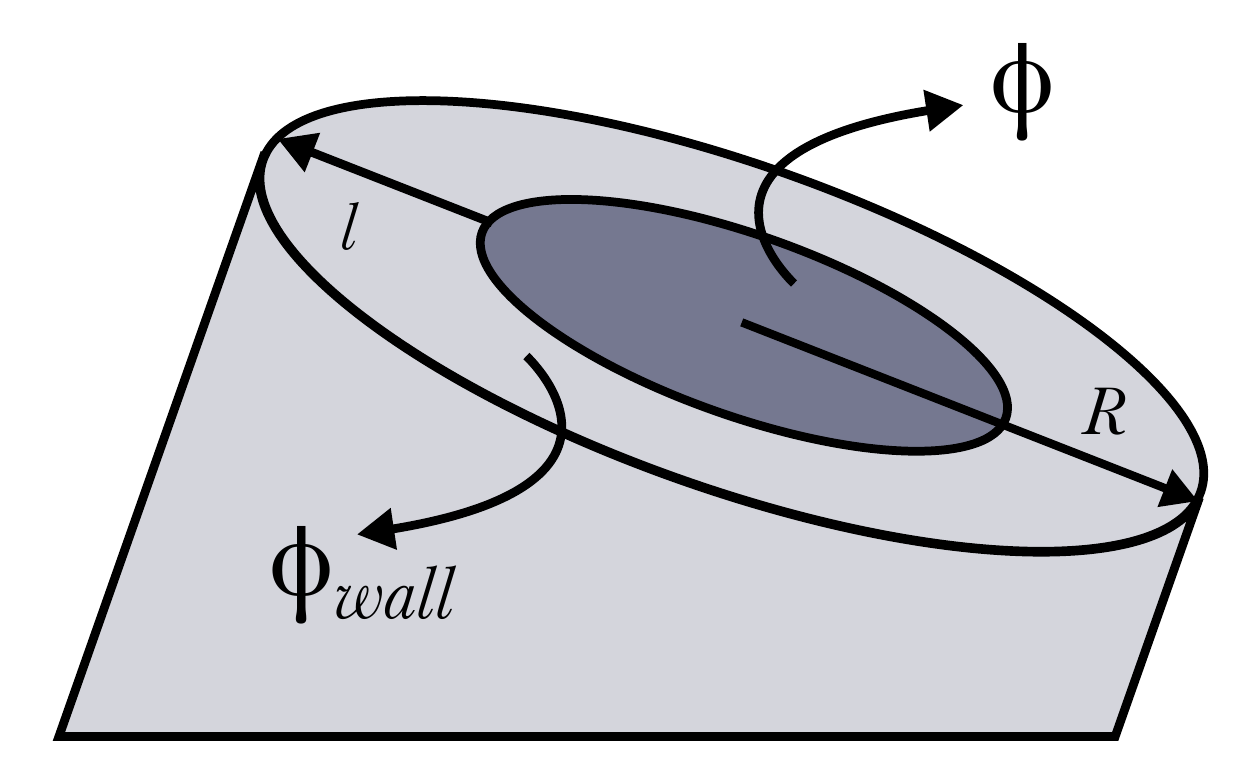}
  \caption{Schematic of particle packing in a cylindrical vessel of radius $R$. The darker shaded area indicates the core region of the packing, where the packing density approaches that of the infinite limit, $\phi$, while wall effects produce a lower density $\phi_{wall}$ in an annulus of width $l$.}
  \label{cyl_diagram}
\end{figure}

In a cylindrical vessel the packing fraction near the wall is slightly lower due to excess voids, while in the interior the packing fraction quickly approaches that of an infinite system. The cross-over between these two regions can be tracked by x-ray tomography\cite{Suz:08} and, as shown by extensive simulations,\cite{Des:09} the radial decay of excess voids from the wall inward is generally well approximated by modeling it as two coaxial regions of slightly different packing density,  $\phi_{wall}$ inside the boundary layer of width $l$ adjacent to the wall and $\phi$ in the central core region (Fig.~\ref{cyl_diagram}). The experimentally measured, net packing density in a cylinder of inner radius $R$ can then be expressed as\cite{Mis:14}

\begin{equation}
\phi_{exp} = \phi_{wall}\left(1-\left(1-\frac{\lambda}{R}l\right)^2\right)+\phi\left(1-\frac{\lambda}{R}l\right)^2
\end{equation}

The normalization factor $\lambda$ is used to express the vessel geometry in terms of a characteristic particle-scale feature so that results from different particle types can be compared on the same plot. It is here set to $\lambda$ = $a$ = 2mm, the radius of the overlapping spheres composing the trimer, which is also the average radius of the apherical ellipsoids. The boundary layer width $l$ is measured in units of $\lambda$.  We used Eq. 1 to fit the experimental results and derive the packing density $\phi$ for an infinite system by extrapolating to the limit $\nicefrac{a}{R} \rightarrow$ 0.

\begin{figure}[h]
\centering
  \includegraphics[height=6cm]{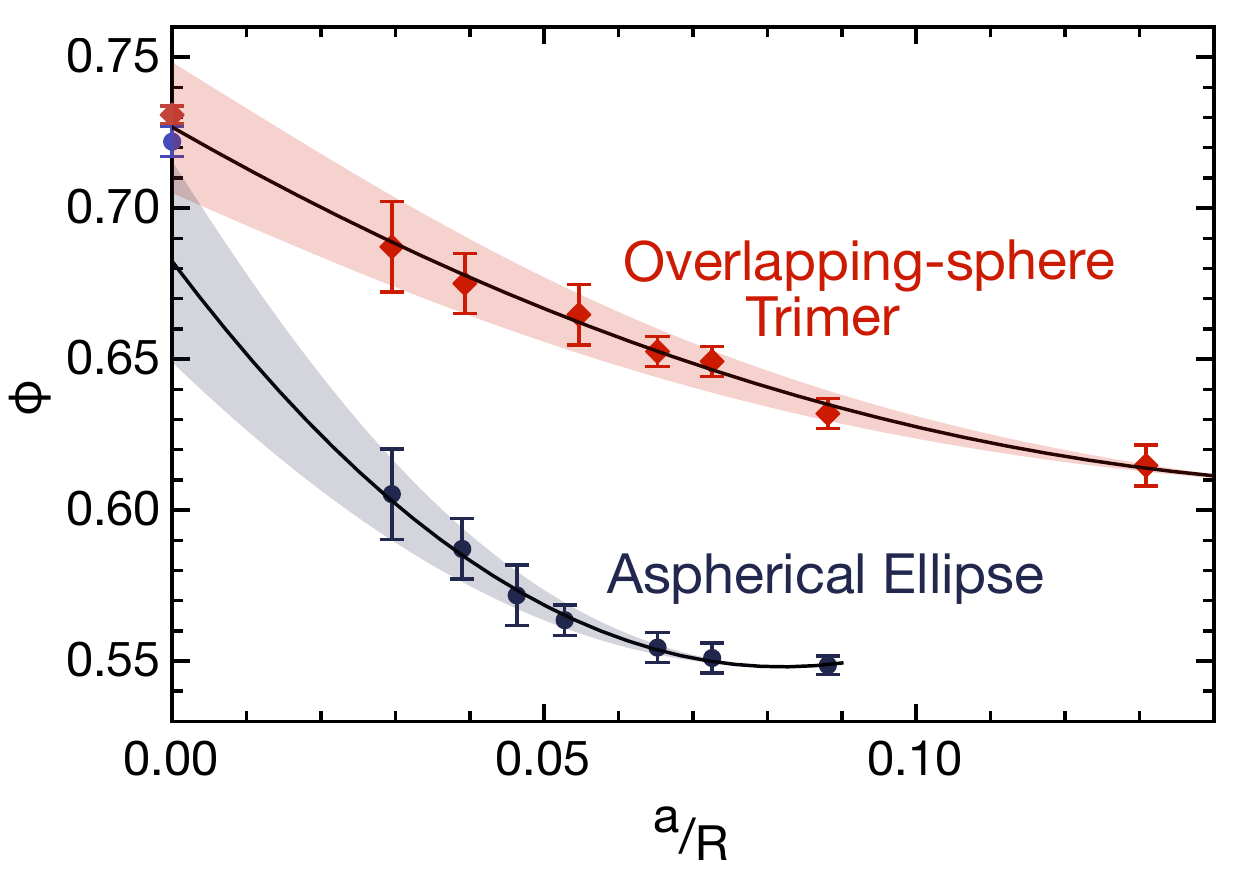}
  \caption{Packing fraction for tapped cylinders, as a function of inverse container radius. The simulation results are plotted at $\nicefrac{a}{R}=0$.}
  \label{exp_plot}
\end{figure}

Our results for $\phi_{exp}$ as a function of inverse system size $\nicefrac{a}{R}$ are plotted in Fig. \ref{exp_plot}, together with fits to Eq. 1. The shaded regions indicate the range of core packing fractions $\phi$ that lie within each fit's uncertainty. The best fit to the experimental data for the overlapping-sphere trimer gives $\phi$ = 0.727$\pm$0.022, which is within error of the packing fraction predicted by our simulation, 0.729$\pm$0.003, shown on the same plot at $\nicefrac{a}{R}$ = 0. From the fit, we extract a boundary layer width for the trimer of approximately 3 individual sphere diameters or about 2.1 particle side lengths.

For the ellipsoids, the best fit gives an infinite system packing fraction $\phi$ = 0.682$\pm$0.033, less than that found by our simulations, 0.722$\pm$0.005, and noticeably lower than the extrapolated densities near $\sim$ 0.74 reported earlier.\cite{Man:05,Don:04} Some of this discrepancy might be due to differences in the preparation protocol for the packings as well as in the experimental procedure. Man et al. followed a somewhat different approach in extrapolating from finite size containers to derive the packing density of the infinite system.\cite{Man:05} Specifically, by assuming that the change in packing density as a function of vessel radius is sufficiently small to make higher order corrections in \nicefrac{a}{R} negligible, Man et al. kept only a linear extrapolation $\phi - \phi_{exp} \sim \nicefrac{a}{R}$.

For our data, it is clear that such a linear relationship would not be sufficient to describe the observed behavior for either the trimer or the aspherical ellipsoid. The steep decay in $\phi$ for the ellipsoid as  $\nicefrac{a}{R}$ is increased, i.e., as the container diameter is reduced, indicates a large boundary layer. At 6.1 particle diameters this layer is more than twice as wide as that of the overlapping-sphere trimer. Thus, for the ellipsoid significantly larger containers are required to minimize the influence of the container walls and to allow for linear extrapolation.

The large boundary layer of the ellipsoids may also be a reason for the discrepancy between our simulated results and those predicted by extrapolation from our experiment. In Eq. 1, we assumed that $\phi_{wall}$ is constant throughout the boundary layer, and that $l$ is not affected by the curvature of the walls. For a layer width of 6 particles this may not fully capture the behavior we see in Fig.~\ref{exp_plot}, and there may indeed be some radial dependence of $\phi_{wall}$ that would need to be accounted for in order to extrapolate to the limit $\nicefrac{a}{R} \rightarrow$ 0 more accurately.

However, irrespective of any extrapolation, a striking outcome from Fig. \ref{exp_plot} is that the new trimer particle identified by the evolutionary algorithm outpaces the ellipsoid by packing more densely at all finite values of \nicefrac{a}{R}, an effect most pronounced for smaller containers. For applications where the target is to achieve a high packing fraction this can be a significant practical advantage.

The mechanism by which this dense packing is accomplished appears to be quite subtle, and so far we have been unable to link it to structural features in the packed aggregate. This is in contrast to the case of the non-overlapping-sphere trimer (Fig. \ref{diff_shapes}a), for which the size of the two small `ears' and their bond angle could be associated with efficiently filling interstitial voids between the larger central spheres.\cite{Mis:14} In simulations of tetrahedral particles comprised of four overlapping spheres, Az\'ema et al. find non-monotonic behavior of $\phi$ with degree of corrugation, where the highest $\phi \sim$ 0.7 occurs for intermediate corrugation,\cite{Aze:13} presumably due to preferential interlocking. The fact that our simulations find very similar packing density (within numerical accuracy) for the particle composed of just three overlapping spheres and the smoothed out triangular version (Figs. \ref{diff_shapes}c \& d) suggests that contacting trimers do not exploit corrugation or dimples to align.

Their planar structure, however, could induce some alignment that might lead to (very local) stacking. We can see hints of this in $g$($r$) (Fig. \ref{g_of_r}).  The small residual dimples in the trimer, while presumably not driving alignment,  nevertheless slightly amplify the likelihood of certain characteristic configurations between neighboring particles, and the resulting peaks in the left shoulder of $g$($r$), indicated by the circles in Fig. \ref{g_of_r}, allow us to pick out the corresponding particle-particle orientations. The closest possible distance between the centers of  two neighboring trimers, 2$a$ or one constituent sphere diameter, requires precise on-top stacking and thus is relatively rare; however, a number of other combinations, such as one 90-degree flipped trimer abutting any one of the sides of the other are more common. Together with the most likely arrangement, a side-by-side configuration of two trimers, they give rise to a large and broad first peak around $r = 2.5a$.

\section{Conclusions}

Approximating arbitrary particle shapes by sets of overlapping spheres, we used an evolutionary algorithm to identify sphere configurations that maximize the particle density $\phi$ for random packings. We found a class of planar triangular particles that pack exceptionally well even in finite-size containers, with $\phi \sim$ 0.73 far outpacing the random packing density achieved by the best-performing particles comprised of non-overlapping spheres and almost approaching the density of crystalline sphere packings.  Specifically, we showed that under identical experimental conditions these particles outperform aspherical ellipsoids, which have one of the highest $\phi$ values reported previously.
 

To the extent that the simulation of the virtual experiments can  account for temperature, the methodology introduced here could be readily extended to optimizing a wide range of other systems, where shaking or thermal energy introduces random particle motion. While we focused on a given, fixed sample preparation protocol, the same type of approach could furthermore be used to optimize the protocol, for example by tuning a sequence of `annealing' steps. This could be done either for a given particle type or together with finding the most suitable particle for a specified system. This framework thus has a compelling ability to tune desired aggregate response in a wide range of applications.

\section*{Acknowledgements}We thank Marc Miskin, Kieran Murphy, Tom Witten, Juan de Pablo, and Victor Lee for insightful discussions. All particle images were generated with Ovito (http://ovito.org/). \cite{Stu:10} This work was supported by the National Science Foundation through CBET---1334426. We acknowledge additional support through Grant No. 70NANB14H012 from the U.S. Department of Commerce, National Institute of Standards and Technology as part of the Center for Hierarchical Material Design (CHiMaD).\\




\bibliography{optim} 
\bibliographystyle{rsc} 

\end{document}